\documentclass[12pt,nofootinbib]{revtex4-1}
\usepackage{graphicx}
\usepackage{cancel}
\usepackage{textcomp}
\usepackage{amsmath, amssymb, amsfonts, latexsym, epsfig}
\usepackage{mathrsfs}
\usepackage{axodraw4j}
\usepackage{pstricks}
\usepackage{bm}
\usepackage{times}
\usepackage{epsfig}
\usepackage{color}
\usepackage{multirow}

\def\beq{\begin{equation}}
\def\eeq{\end{equation}}
\def\be{\begin{equation}}
\def\ee{\end{equation}}
\def\bea{\begin{eqnarray}}
\def\eea{\end{eqnarray}}
\def\to{\rightarrow}

\begin{document}
\title{Is polarization effect visible in leptonic SUSY searches?}
\author{Kai Wang~$^{a}$}
\author{Liucheng Wang~$^{a,b}$}
\author{Tao Xu~$^{a}$}
\author{Liangliang Zhang~$^{a,c}$}
\affiliation{$^{a}$ Zhejiang Institute of Modern Physics and Department of Physics, Zhejiang University, Hangzhou, Zhejiang 310027, CHINA\\
$^{b}$  Bartol Research Institute, Department of Physics and Astronomy,
University of Delaware, Newark, DE 19716, USA\\
$^{c}$ School of Physics, Nankai University, Tianjin 300071, CHINA}

\begin{abstract}
On-shell effective theory approach has been widely used in search of various supersymmetric signals, in particular, gluino/squark pairs with long cascade decay chains in which complete matrix element calculations may encounter over-20 dimensional integrations. On the other hand, leptons from polarized chargino decays may show significant boost or anti-boost effect in some scenarios and simulation without polarization information may underestimate or overestimate the lepton $p_{T}$ cut efficiencies in the first place. We study the polarization effects in supersymmetry searches of multi-jets plus leptons final states. We find it justifiable for first two generations to only use on-shell effective theories. While for measurements related to third generation squarks, for instance,  polarization effect of charginos from stop may reduce the lepton $p_{T}$ cut efficiencies in cross section measurements by 25\% when slepton contributions dominate in chargino decay or $W$ are on-shell.
The signal is then underestimated if only on-shell effective theory approach is taken in simulation of signal and the real bound on squark/gluino should be more stringent.   
\end{abstract}

\maketitle

\section{Introduction}

With two years' running, Large Hadron Collider (LHC) at CERN has accumulated  data of 30~fb$^{-1}$ 
integrated luminosity. In the early stage, most of the discoveries are kinematics dominated.
On-shell effective theories (OSET) which characterize hadron collider data 
in terms of masses, production cross sections, and decay modes 
of candidate new particles plays important role in new physics searches~\cite{ArkaniHamed:2007fw}. In principle, if there exists the precision predictions of production cross section and  decay mode for given mass of candidate new particle, one can also obtain information on its spin. However, direct confirmation of spin state measurement only comes from measurement of angular correlation. With only about 28~fb$^{-1}$, a Higgs boson has been discovered with over 7-$\sigma$ significance via measurement of invariant masses  for four-lepton final states. Both ATLAS
and CMS collaborations have found a resonance of four-lepton with invariant mass of 125~GeV and the reconstructed di-photon invariant mass also peaks at the same place. 
The di-photon final state has excluded the boson to be spin-one state based on argument from Landau-Yang theorem. The further analysis of spin/parity measurement of the boson based on data on the angular correlation in four-lepton channel is very compatible with the  scalar boson expectations of $0^{+}$. The data disfavors the $2^{+}$ hypothesis with a Confidence Levels value of 0.6\% \cite{today}.  
Whether the boson is the standard model (SM) Higgs then requires additional measurements of its
Yukawa couplings to the SM fermions. 

On the other hand, the transverse momentum distribution $d\sigma/d p_{T}$  is 
not only a result of mass spectrum but 
also dependent on polarization of spin-1 or spin-$1/2$ particles,
while the measurement of cross section significantly depend on the simulated cut efficiency. 
Consequently, polarization effect may in principle come in for the very early stage measurement of cross section. A full matrix element simulation automatically contains all spin and polarization information. However, in many cases, searches may involve multi-body final states which correspond multi-body phase-space integration. A $n$-body phase space 
is $3n-4$ dimensional integration with the additional two dimensions over the two initial parton distributions
so $3n-2$ dimensional integration is then required. 
High dimensional integration is technically extremely challenging. 
Therefore, simulation based on kinematic decay is sometimes inevitable.  
For instance, Gluino cascade decay into lepton final state 
\beq
\tilde{g} \to j j  \chi^{\pm}_{1} \to j j \ell^{\pm}\nu_{\ell} \tilde{\chi}^{0}_{1}
\eeq
involves 5 body final states in one chain. Gluino pair production with di-lepton final states
then corresponds to 10 body and the total integration is 28 dimensional integration \cite{TheATLAScollaboration:2013uha}. 
The di-lepton plus jets arising from squark pair or gluino-squark production then corresponds to 8 or 9 body final states
and 22 or 25 dimensional integration. The simulation of these final states are typically done by only kinematics in Pythia.  
In this paper, we study how polarization effect changes the measurement of production cross section in particular. 
We compare the lepton $p_{T}$ cut efficiency between study involving polarization and study with pure kinematic decays
in different scenarios. 

In the next section, we discuss two examples where polarization play important role in measurements. Then we discuss the 
chargino polarization in squark/gluino decay and its leptonic decay distribution in different scenarios, for instance, light sleptons,
off-shell $W$ or on-shell $W$. We then gives the numeric results of comparison before we conclude. 

\section{Polarization Effects}

Polarization effect has been widely studied in LEP era. 
We use two examples to illustrate the polarization effects used in collider physics.

In one-prong decay of $\tau$ lepton
$\tau^{-}\to \pi^{-}\nu_{\tau}$~, the nearly massless $\nu_{\tau}$ is of left-handed helicity 
and pion $\pi^{-}$ is a pseudo-scalar state. In the left-handed $\tau^{-}_{L}$ rest frame, 
neutrino $\nu_{\tau}$ is moving in the direction of $\tau^{-}_{L}$ boosted direction 
and pion is boosted in the opposite direction of $\tau^{-}_{L}$ moving direction.
While for right-handed $\tau^{-}_{R}$, the pion $\pi^{-}$ is boosted in the $\tau^{-}_{R}$ 
moving direction. Therefore, left-handed or right-handed $\tau$ can be clearly distinguished
via the $p_{T}$ measurement of pions \cite{FileviezPerez:2008bj}. 

This feature can be applied to search of new charged 
Higgs state $H^{\pm}$ since $\tau^{\pm}$ from $H^{\pm}$ decay and the leading irreducible background of $W^{\pm}$ decay are in different helicity states. A simple $p_{T}$ cut in charged pion of $\tau$-tag 
can significantly reduce the SM background from $W^{\pm}$ decay. 

\begin{figure}[ht]
\begin{center}
\includegraphics[scale=1,width=8cm]{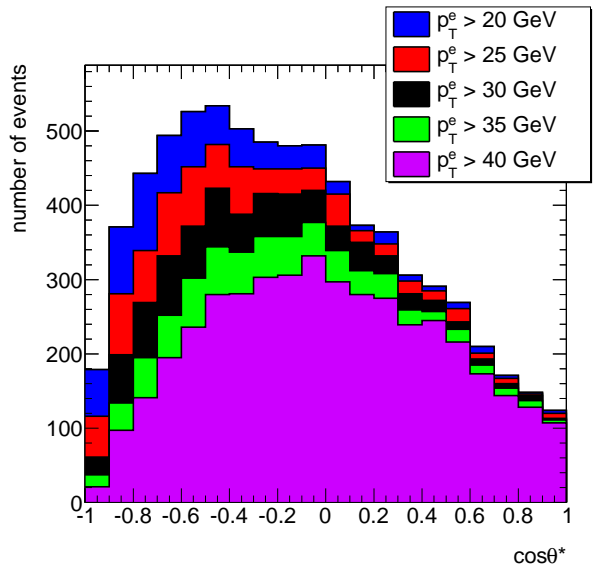}
\end{center}
\caption{$d\Gamma/d\cos\theta$ distribution in SM semi-leptonic $t\bar{t}$ with different $p_{T}$ selection cuts as labelled in plot.}
\label{ttb}
\end{figure}

A second and more similar example is the measurement of $W$-polarization 
in top quark decay $t\to b W^{+}$ which plays role as test of 
the Higgs mechanism in SM \cite{Kane:1991bg}. In so-called ``Higgs'' mechanism, the
Goldstone degree of freedom becomes the longitudinal polarization
of $W$ boson, $\epsilon^{0}$. Since top quark is the heaviest particle
which acquire mass through electroweak symmetry breaking (EWSB), 
top quark couples to the Goldstone boson strongly which
results in $m_{t}/m_{W}$ enhancement in $\epsilon_{\mu}^{0*}\bar{u}_{b}P_{L}\gamma^{\mu}u_{t}\propto m_{t}/m_{W}$.
The $W$ bosons from top quark decay can be produced longitudinal, left-handed
and right-handed, with the helicity fractions $F_{0}$, $F_{L}$ and
$F_{R}$ respectively. Due to $m_{t}/m_{W}$ enhancement, one has
$F_{0}=\frac{(m_{t}/m_{W})^{2}}{(m_{t}/m_{W})^{2}+2}=70\%$, $F_{L}=\frac{2}{(m_{t}/m_{W})^{2}+2}=30\%$
and $F_{R}=0$ in the limit $m_{b}=0$. Theoretically, the precision
predictions of helicity fractions are obtained by next-to-next-to-leading
order (NNLO) pQCD calculations. Given the full reconstruction of semi-leptonic $t\bar{t}$ 
system, polarizations can be measured by measuring the angle between 
the lepton and $W$ in the $W$-rest frame.  
In left-handed polarized $W^{+}$ $W^{+}\to (e^{+})_{R} (\nu_{e})_{L}$,
the lepton is boosted in the opposite moving direction of $W^{+}$ with
\beq
\mid{\cal M}(W^{+}_{L}\to e^{+}\nu_{e})\mid^{2} \propto (1-\cos\theta)^{2}
\eeq
where $\cos\theta$ is defined as the angle between electron and the opposite direction
of $b$-jet in the $W$-rest frame. The measured lepton momentum in the lab frame are then softer than
the ones in $W$ rest frame and softer leptons may not pass the lepton $p_{T}$ cut.
Therefore, the portion of $F_{L}$ measurement also depends on
the lepton $p_{T}$ cut. Figure \ref{ttb} shows the $d\Gamma/d\cos\theta$ distribution in SM 
semi-leptonic $t\bar{t}$ with different $p_{T}$ selection cuts as labelled in plot
from $p_{T}: 20-40$~GeV.

\section{Polarized Chargino}

In this section, we focus on the multi-body final states with leptons due to gluino or squark cascade decays. 
Similar to $Z$ decay, Leptonic decay branching ratio (BR) of heavy neutralino $\tilde{\chi}^{0}_{2}$ is typically much smaller 
than its of $\tilde{\chi}^{\pm}_{1}$. In addition, BR of $\tilde{\chi}^{0}_{2}$ final state in squark decay is also 
smaller than squark decay BR into charginos $\tilde{\chi}^{\pm}_{1}$. Therefore, leptons largely arise from 
chargino decay $\tilde{\chi}^{\pm}_{i} \to \ell^{\pm} \nu\tilde{\chi}^{0}_{1}$ and in this paper, we focus on studying the
effect of $\tilde{\chi}^{\pm}_{i} \to \ell^{\pm} \nu\tilde{\chi}^{0}_{1}$ of polarized $\tilde{\chi}^{\pm}_{i}$. 

\begin{figure}[ht]
\begin{center}
\includegraphics[scale=1,width=6cm]{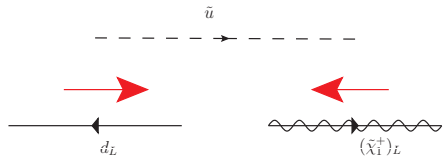}
\end{center}
\caption{Spin correlation in squark decaying into charging, left-handed polarized $\tilde{\chi}^{+}_{1}$ }
\label{usquarkdecay}
\end{figure}

Chargino states are mixture of both Wino and Higgsino which is determined by $M_{2}$ and $\mu$ in the mixing matrix. 
We first discuss Wino chargino for general squark decay. Decay BR into Higgsino of squark is only significant for third generations
and we discuss later as a special case.  Wino is a super-partner of $SU(2)_{L}$ weak gauge boson $W$ and the squark-quark-wino vertex 
\beq
{\cal L}=[-g V_{11}\tilde{u}_{L}]\bar{d}P_{R}\tilde{\chi}^{+c}_{1}
\label{wino}
\eeq
corresponds to weak interaction where only left-handed quark or right-handed anti-quark participates the interaction. 
As shown in Fig. \ref{usquarkdecay}, left-handed up-type scalar quark $\tilde{u}_{L}$ decay into left-handed $b$-quark,
which results in $\tilde{\chi}^{+}_{1}$ to be only left-handed. 
Since the scalar propagator does not carry any spin information, the two fermion lines between
the scalar mediator have no correlation which does not depend on whether the scalar is on-shell or
off-shell. If squarks are heavier than gluino, gluino three body decay
\beq
\tilde{g}\to \bar{u} d \tilde{\chi}^{+}_{1}~,
\eeq
the polarization is identical to the case when squarks are on-shell and hence, the discussion on
chargino polarization stays the same for off-shell squarks.   
When the chargino is Wino, for all squarks and anti-squarks, one can write down the similar relations as 
\bea
\tilde{u}_{L} &\to & d_{L} (\tilde{\chi}^{+}_{1})_{L},~~~~
\tilde{u}^{*}_{L} \to  (\bar{d})_{R} (\tilde{\chi}^{-}_{1})_{R}\nonumber\\
\tilde{d}_{L} &\to & u_{L} (\tilde{\chi}^{-}_{1})_{L},~~~~
\tilde{d}^{*}_{L} \to  (\bar{u})_{R} (\tilde{\chi}^{+}_{1})_{R}~.
\eea 
On the other hand, for first two generations, scalar quarks decay into light quark states which hadronize immediately at $1/\Lambda_{QCD}\sim 10^{-24}$~s. It is impossible to distinguish whether the light quark jet
is from up-type or down-type quarks. The identical final states then leads to simultaneous measurements of 
first two generation scalar quarks. 
However, decay final state of stop is clearly identifiable \cite{Low:2013aza}.  
\beq
\tilde{t}_{L} \to b_{L}(\tilde{\chi}^{+}_{1})_{L},~~ \tilde{t}^{*}_{L} \to (\bar{b})_{R}(\tilde{\chi}^{-}_{1})_{R}
\eeq
The argument then applies to the stop search. When top $A$-term $A_{t}$ is large which is preferred by MSSM parameter space for $m_{h}=125$~GeV, stops are a mixture of left-handed and right-handed stop states. The stop $\tilde{t}_{1}$, $\tilde{t}_{2}$ can then decay into 
\beq
\tilde{t}_{i}\to t\tilde{\chi}^{0}_{1}, b\tilde{\chi}^{+}_{1}
\eeq
where  $b\tilde{\chi}^{+}_{1}$ final state only comes form left-handed stop state $\tilde{t}_{L}$. When $\tilde{t}_{i}$ has significant $\tilde{t}_{L}$ portion, for instance 50\%, $b\tilde{\chi}^{+}_{1}$ will usually take 70\% of $\tilde{t}_{i}$ decay. When $t$ is not on-shell where
mass difference $M_{\tilde{t}_{i}}-M_{\tilde{\chi}^{0}_{1}} < m_{t}$,  the multi-body phase space will further suppress the decay of $\tilde{t}_{i}\to t^{*}\tilde{\chi}^{0}_{1}$. However,
in this degenerate spectrum limit, the boost of chargino is small and the polarization effect
is then limited.  

Direct search of Chargino via tri-lepton channels at LHC \cite{ATLAS:2013rla}
has pushed the chargino mass to be over 350~GeV when $\tilde{\chi}^{0}_{1}$ is massless but the bound is certainly weaker when the mass difference between chargino and lightest neutralino is smaller. We use two un-excluded benchmark points to illustrate the feature. One benchmark is the light chargino with nearly degenerate spectrum where 
$M_{\tilde{\chi}^{+}_{1}}-M_{\tilde{\chi}^{0}_{1}} \simeq 50$~GeV with $M_{\tilde{\chi}^{+}_{1}}\simeq 250~$GeV.
The second benchmark is $M_{\tilde{\chi}^{+}_{1}}\simeq 285$~GeV and the same Bino mass as the first one
so that $M_{\tilde{\chi}^{+}_{1}}-M_{\tilde{\chi}^{0}_{1}}> m_{W}$ and $W$ is on-shell. 

\begin{figure}[ht]
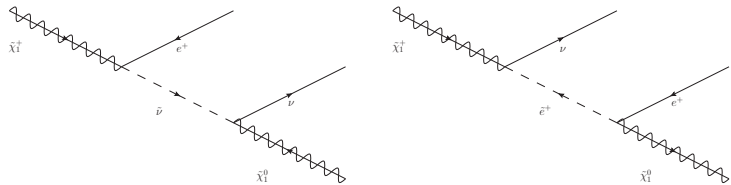

\begin{center}
\includegraphics[scale=1,width=5cm]{decaysneutrino.epsi}
\includegraphics[scale=1,width=5cm]{decayselectron.epsi}
\end{center}
\caption{$\tilde{\chi}^{+}_{1}\to e^{+}\nu_{e}\tilde{\chi}^{0}_{1}$ via sleptons}
\label{decayslepton}
\end{figure}

For the first benchmark point of chargino three body decay 
\beq
\tilde{\chi}^{+}_{1}\to\tilde{\chi}^{0}_{1}+e^{+} +\nu_{e},
\eeq
can be realized by three processes shown in Fig.\ref{decayslepton} with slepton mediation and Fig.\ref{decayw} of $W$ mediation.

\begin{figure}[ht]
\begin{center}
\includegraphics[scale=1,width=5cm]{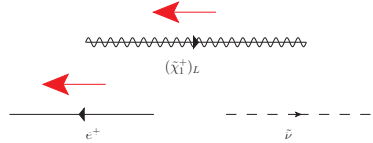}
\end{center}
\caption{Spin correlation in left-handed $\tilde{\chi}^{+}_{1}$ decay into sneutrino $\tilde{\nu}$. }
\label{sneutrinospin}
\end{figure}

In sneutrino mediation as in Fig.\ref{decayslepton}-a, sneutrino propagator does not carry any spin so the spin
direction of positron is identical to the initial chargino spin. Only right-handed positron participates in weak interaction and therefore, positron moves in the opposite direction of the initial chargino as shown in Fig.\ref{sneutrinospin}
 
In selectron mediation as in Fig.\ref{decayslepton}-b, electron decay from spin-zero selectron  
does not correlate with the initial polarization of chargino in both on-shell and off-shell selectron case. 
The electron is therefore completely universal distributed in space. 

For $W$-mediation, 
\begin{figure}[ht]
\begin{center}
\includegraphics[scale=1,width=5cm]{decayw.epsi}
\end{center}
\caption{$\tilde{\chi}^{+}_{1}\to e^{+}\nu_{e}\tilde{\chi}^{0}_{1}$ via $W^{+}$}
\label{decayw}
\end{figure}
the situation then depends on whether $W$ is on-shell or off-shell.  
For the first benchmark point, $W$ is off-shell. It is known that there
exist four polarizations in the off-shell $W$ case including additional
scalar contribution. However, in the nearly massless final state in the
leptonic decay case, scalar contribution and the scalar-longitudinal 
interference contribution becomes negligible \cite{Groote:2013xt}. 

\begin{figure}[ht]
\begin{center}
\includegraphics[scale=1,width=7cm]{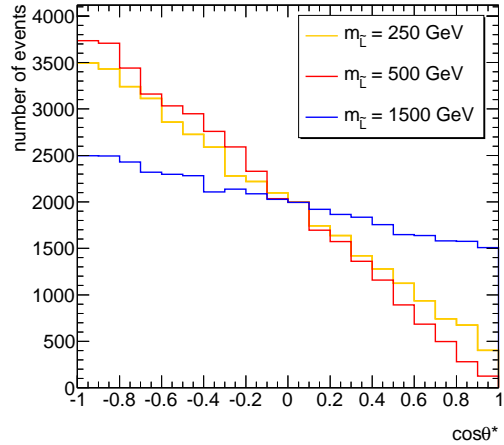}
\end{center}
\caption{Normalized distribution $1/\Gamma d\Gamma /d\cos\theta$ where $\theta$ is the angle between charged lepton and chargino moving direction in reconstructed $\chi^{+}_{1}$ frame for $\Delta M=50$~GeV with different slepton masses $M_{\tilde{\ell}}$.}
\label{costheta}
\end{figure}

The contributions also destructively interfere with each other. 
Figure \ref{costheta} shows the angle $\theta$ between the positron and chargino in the reconstructed chargino rest
frame for three choices of slepton masses. For simplicity, we assume $M_{\tilde{\ell}}=M_{\tilde{e}}$ with negligible lepton 
$A$-terms. $M_{2}$ is fixed at 250~GeV while $M_{1}$ is taken to be 200~GeV. 
For $M_{\tilde{\ell}}=250$ and 500~GeV, the sneutrino and selectron contributions are both significant. 
Since Bino is SM singlet, $W$ contribution is only realized through Wino-Bino mixing which is small. 
Therefore, we find the slepton contribution does not decouple even when $M_{\tilde{\ell}}=1.5$~TeV. 

\begin{figure}[ht]
\begin{center}
\includegraphics[scale=1,width=7cm]{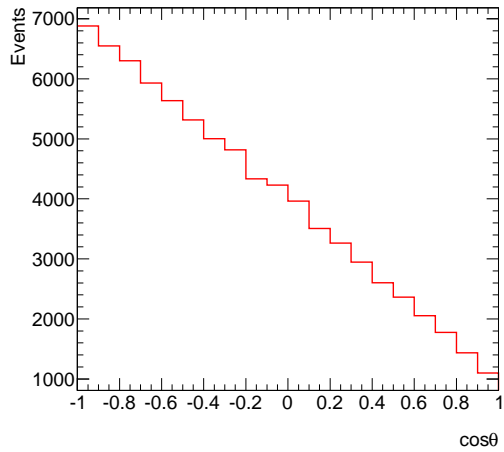}
\end{center}
\caption{Normalized distribution $1/\Gamma d\Gamma /d\cos\theta$ where $\theta$ is the angle between charged lepton and chargino moving direction in reconstructed $\chi^{+}_{1}$ frame for on-shell $W$ case.}
\label{costheta2}
\end{figure}

When $W$ is on-shell for the second benchmark point, the two-body decay
contribution completely takes over and we plot the $\cos\theta$ for on-shell $W$ case in Fig.\ref{costheta2}.
The positron is always moving in the opposite direction of the left-handed polarized chargino.

Higgsino couples to squarks via Yukawa couplings and hence squarks of first two generations
mostly decay into wino states. The third generation squarks, stop or sbottom of relatively large $\tan\beta$,
can dominantly decay into Higgsino states. Equation \ref{stop} gives the interaction between stop and charginos
\beq
{\cal L}_{b\tilde{t}\tilde{\chi}^{\pm}}=[-g V_{i1}\tilde{t}_{L}+y_{t}V_{i2} \tilde{t}_{R}](\bar{b}P_{R}\tilde{\chi}^{+c}_{i}) + y_{b}U^{*}_{i2} \tilde{t}
_{L}(\bar{b}P_{L}\tilde{\chi}^{+c}_{i})~,
\label{stop}
\eeq
where $U_{ij}$ and $V_{ij}$ are standard chargino mixing matrices. If $\tilde{\chi}^{\pm}_{1}$ is 
dominated by Higgsino component and $y_{t}$ is much larger then $y_{b}$, $\tilde{\chi}^{\pm}_{1}$
polarization in the decay is identical to the case of wino $\tilde{\chi}^{\pm}_{1}$. Only $\tan\beta\sim 30$ or greater,
bottom Yukawa $y_{b}$ is then close to $y_{t}$ and $\tilde{\chi}^{\pm}_{1}$ polarization then has 
both significantly left-handed and right-handed portions. However, supersymmetric scenario with
large $\tan\beta$ is severely constrained by rare decay process like $B_{s}\to \mu^{+}\mu^{-}$. For small $\tan\beta$ cases,
Higgsino behavior is then similar to the Wino. 
On the other hand, since the electron/muon Yukawa couplings are negligible Higgsino decay into lepton is only an effect of on-shell
or off-shell $W$ which  has been discussed  in the last part of Wino case. 

\section{Lepton in Cascade Decay Chain}

As discussed in previous section, 
polarization of chargino can be distinguished in third generations squarks decays since the up-type and down-type squarks
have different and identifiable final states. Chargino from sbottom decay is associated with top quark which complicates the
studies of final states. Therefore,  in this section, we focus on chargino from stop decay to illustrate the polarization feature. 
In this case, only left-handed $\tilde{\chi}^{+}_{1}$ can arise from the stop decay
\beq
\tilde{t}_{i}\to b_{L} (\tilde{\chi}^{+}_{1})_{L}~,
\eeq
where left-handed polarized chargino decaying into lepton final states may show significant anti-boost effect. 
Consequently, it may make the leptons softer in the lab frame than the ones in chargino rest frame. Since it's a boosted
effect, the effect is visible only when chargino has significant kinematic boost. If the chargino is produced nearly at rest which corresponds
to the scenario with  small mass difference between chargino and stop/gluino, the effect is then tiny.  On the other hand, if
the mass difference between chargino and stop/gluino is too large, the chargino is highly boosted in the system and lepton momentum in the lower end are then completely dominated by the boost effect. Since we are only looked at the consequence of lepton $p_{T}$ cut
which only affects softer leptons, the effect is only significant in some intermediate mass range. 

\begin{figure}[h]
\begin{center}
\includegraphics[scale=1,width=8cm]{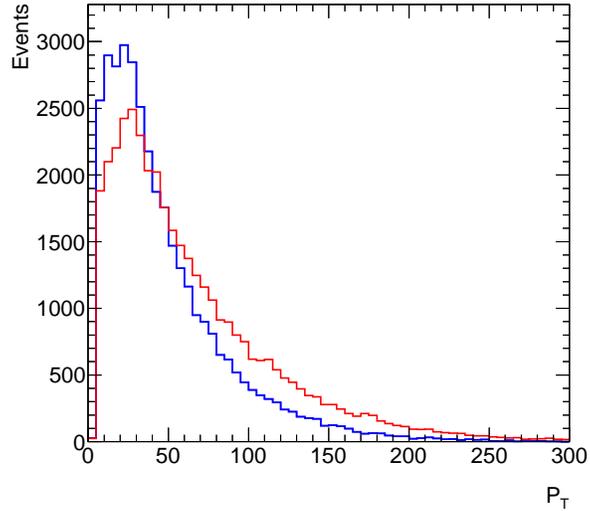}
\end{center}
\caption{Lepton $p_{T}$ distribution of chargino decay $\tilde{\chi}^{\pm}_{1}\to\ell^{\pm}\nu\tilde{\chi}^{0}_{1}$ in comparison of full matrix element method vs. kinematic method. Blue line is the full matrix element result from MadEvent while the red line is the kinematic decay from Pythia.}
\label{pt}
\end{figure}

Figure \ref{pt} shows the lepton $p_{T}$ distribution of chargino decay $\tilde{\chi}^{\pm}_{1}\to\ell^{\pm}\nu\tilde{\chi}^{0}_{1}$ in comparison of full matrix element method vs. kinematic method for $M_{\tilde{t}}=1.3$~TeV and $M_{\tilde{\chi}^{\pm}_{1}}=150$~GeV. Blue line is the full matrix element result from MadEvent\cite{Alwall:2011uj} while the red line is the kinematic decay from Pythia\cite{Sjostrand:2006za} which shows the full matrix element result is significant softer than the result from kinematic decay.

The lepton $p_{T}$ cut survival probabilities for two benchmark points with the same chargino mass $M_{\tilde{\chi}^{\pm}_{1}}=$~GeV but $M_{\tilde{t}}=1.3$ and 1.5~TeV are listed for 
different $p_{T}$ cut respectively in Table \ref{table}. 
\begin{table}[h]
\begin{center}
\begin{tabular}{|c|c|| c |c| c| c|}
\hline
$M_{\tilde{t}}$ & Category &  $p_{T} >20$~GeV  & $p_{T} >25$~GeV &  $p_{T} >30$~GeV\\
\hline
\multirow{2}{*}{$1.3$~TeV} & Polarized & 52\% & 46\% & 40\% \\
& Kinematic  & 64\% & 59\% & 54\% \\
\hline
\multirow{2}{*}{$1.5$~TeV} & Polarized   & 54\% &  48\% & 44\% \\
& Kinematic  & 65\% & 61\% & 57\% \\
\hline
\end{tabular}
\end{center}
\label{table}
\caption{Lepton $p_{T}$ cut survival probability for benchmark points $M_{\tilde{t}}=1.3$ and 1.5~TeV. Category named polarized is for
situation with polarization taken into account while  Kinematic stands for the kinematic decay only cases. }
\end{table}%
The reduction of efficiencies is about 25\% for both cases in Table \ref{table}.
Therefore, if one only use on-shell effective field theory approach for this search, the signal is actually under-estimated.

\section{Conclusions}
We study the polarization effects in supersymmetry searches of multi-jets plus leptons final states. We find it justifiable for first two generations to only use on-shell effective theories. While for measurements related to third generation squarks, for instance,  polarization effect of charginos from stop may reduce the lepton $p_{T}$ cut efficiencies in cross section measurements by 25\% when slepton contributions dominate in chargino decay or $W$ are on-shell.
The signal is then underestimated if only on-shell effective theory approach is taken in simulation of signal and the real bound on squark/gluino should be more stringent.   

\section*{Acknowledgement}
The project is supported in part, by the Zhejiang University Fundamental Research Funds for the Central Universities (2011QNA3017) and the National Science Foundation of China (11245002,11275168, ).


\begin{thebibliography}{References}


\bibitem{ArkaniHamed:2007fw} 
  N.~Arkani-Hamed, P.~Schuster, N.~Toro, J.~Thaler, L.~-T.~Wang, B.~Knuteson and S.~Mrenna,
  hep-ph/0703088 [HEP-PH].




\bibitem{today}
  G.~Aad {\it et al.}  [ATLAS Collaboration],
son with the ATLAS detector at the LHC,''
  Phys.\ Lett.\ B
  [arXiv:1207.7214 [hep-ex]].
  S.~Chatrchyan {\it et al.}  [CMS Collaboration],
the LHC,''
  Phys.\ Lett.\ B
  [arXiv:1207.7235 [hep-ex]].

\bibitem{TheATLAScollaboration:2013uha} 
  The ATLAS collaboration,
  ATLAS-CONF-2013-062.
  J.~F.~Hirschauer [CMS Collaboration],
  PoS ICHEP {\bf 2012}, 177 (2013).
  
  

\bibitem{FileviezPerez:2008bj} 
  Q.~-H.~Cao, S.~Kanemura and C.~P.~Yuan,
  Phys.\ Rev.\ D {\bf 69}, 075008 (2004)
  [hep-ph/0311083].
  P.~Fileviez Perez, H.~H.~Patel, M.~.J.~Ramsey-Musolf and K.~Wang,
  Phys.\ Rev.\ D {\bf 79}, 055024 (2009)
  [arXiv:0811.3957 [hep-ph]].
  P.~Fileviez Perez, T.~Han, G.~-Y.~Huang, T.~Li and K.~Wang,
  Phys.\ Rev.\ D {\bf 78}, 071301 (2008)
  [arXiv:0803.3450 [hep-ph]].

\bibitem{Kane:1991bg} 
  G.~L.~Kane, G.~A.~Ladinsky and C.~P.~Yuan,
  Phys.\ Rev.\ D {\bf 45}, 124 (1992).
  A.~Czarnecki, J.~G.~Korner and J.~H.~Piclum,
  Phys.\ Rev.\ D {\bf 81}, 111503 (2010)
  [arXiv:1005.2625 [hep-ph]].
  X.~-Q.~Li, Z.~-G.~Si, K.~Wang, L.~Wang, L.~Zhang and G.~Zhu,
  arXiv:1311.6874 [hep-ph].
\bibitem{Low:2013aza} 
  I.~Low,
  Phys.\ Rev.\ D {\bf 88}, 095018 (2013)
  [arXiv:1304.0491 [hep-ph]].
\bibitem{ATLAS:2013rla} 
  [ATLAS Collaboration],
  ATLAS-CONF-2013-035.
  
  
  
\bibitem{Groote:2013xt} 
  S.~Groote, J.~G.~Korner and P.~Tuvike,
  Eur.\ Phys.\ J.\ C {\bf 73}, 2454 (2013)
  [arXiv:1301.0881 [hep-ph]].
  
\bibitem{Alwall:2011uj} 
  J.~Alwall, M.~Herquet, F.~Maltoni, O.~Mattelaer and T.~Stelzer,
  JHEP {\bf 1106}, 128 (2011)
  [arXiv:1106.0522 [hep-ph]].



\bibitem{Sjostrand:2006za} 
  T.~Sjostrand, S.~Mrenna and P.~Z.~Skands,
  JHEP {\bf 0605}, 026 (2006)
  [hep-ph/0603175].
\end{thebibliography}
\end{document}